\documentclass{osa-article}

\journal{osac}

\articletype{Research Article}

\usepackage{tikz}
\usepackage{pgfplots}
\usetikzlibrary{arrows,positioning}
\usetikzlibrary{shapes.geometric, arrows}
\usetikzlibrary{positioning,fit,calc}
\usepackage{graphicx}
\usetikzlibrary{patterns,svg.path}
\usepgfplotslibrary{fillbetween}
\usetikzlibrary{intersections}
\usepackage[caption=false, font=normalsize, labelfont=sf, textfont=sf]{subfig}
\usetikzlibrary{calc,trees,positioning,arrows,chains,shapes.geometric,  decorations.pathreplacing,decorations.pathmorphing,shapes,  matrix,shapes.symbols}
\usetikzlibrary{matrix, arrows.meta, shapes}

\tikzset{
    >=stealth',
    pil/.style={
           ->,
           thick,
           shorten <=2pt,
           shorten >=2pt,}
}

\begin{document}

\title{Stress mirror polishing for future large lightweight mirrors: design using shape optimization.}

\author{Sabri Lemared,\authormark{1} Marc Ferrari,\authormark{1} Christian Du Jeu,\authormark{2} Thibault Dufour,\authormark{2} Nathalie Soulier,\authormark{2} and Emmanuel Hugot\authormark{1}}

\address{\authormark{1}Aix Marseille Univ, CNRS, CNES, LAM, Marseille, France\\
\authormark{2}Thales-SESO, Aix-en-Provence, France}

\email{\authormark{*}sabri.lemared@lam.fr} 



\begin{abstract}
This study proposes a new way to manufacture large lightweight aspherics for space telescopes using Stress Mirror Polishing (SMP). This technique is well known to allow reaching high quality optical surfaces in a minimum time period, thanks to a spherical full-size polishing tool. To obtain the correct surface' aspheric shape, it is necessary to define precisely the thickness distribution of the mirror to be deformed, according to the manufacturing parameters. 
We first introduce active optics and Stress Mirror Polishing techniques, and then, we describe the process to obtain the appropriate thickness mirror distribution allowing to generate the required aspheric shape during polishing phase. Shape optimization procedure using PYTHON programing and NASTRAN optimization solver using Finite Element Model (FEM) is developed and discussed in order to assist this process. The main result of this paper is the ability of the shape optimization process to support SMP technique to generate a peculiar aspherical shape from a spherical optical surface thanks to a thickness distribution reshaping. This paper is primarily focused on a theoretical framework with numerical simulations as the first step before the manufacturing of a demonstrator. This "two-steps" approach was successfully used for previous projects.
\end{abstract}

\section{Introduction}
Stress Mirror Polishing (SMP) is a manufacturing method allowing to obtain the required shape of optical surfaces using deformation and elasticity properties of materials. In this approach, the mirror is deformed from a sphere, as we want to obtain a quasi parabola, to the desired inverse-shape, spherically polished using full-size tools, before to take the correct form when the applied constraints are released. This technique has been developed for forty years on several type of mirrors such as Keck segments \cite{Lubliner:80} or the Spectro-Polarimetric High-contrast Exoplanet Reseach (SPHERE) instrument on the Very Large Telescope (VLT) \cite{refId0}. Contrary to the classical aspherical polishing process where small size tools introduce some high-order spatial frequencies errors on the optical surface, the SMP process uses large tool which gives an extreme surface quality due to the continuous deformation and polishing behaviors. 

Moreover, due to the large area of the polishing tools, SMP process \cite{Izazaga_Perez_SMP} allows a considerable saving in time of manufacturing and is perfectly suited to produce a large number of primary mirror segments in a short time for future space and ground-based observatories such as Thirty Meter Telescope \cite{TMT2010,TMT2012} and LUVOIR \cite{LUVOIR} projects. SMP techniques have also a significant benefit for the manufacturing of large lightweight aspheric primary mirrors for space telescopes, especially as they can be applied on a thin shell mirror and combined with an active support \cite{Burge} or a rigid assembly using a sandwich structure with two facesheets and a Carbon Fiber Reinforced Composite (CFRC) honeycomb\cite{Catanzaro}. This could be perfectly suited for manufacturing of ultra high surface quality primary mirrors, as the ones foreseen in future high-contrast imaging missions for exoplanets detection and characterization, or simply to have a gain in term of manufacturing time for more classical mirrors.

We investigated the possibility to produce such a lightweight aspheric primary mirror for space telescopes using classical elasticity theory combined with new tools. They are provided by Finite Element Analysis (FEA), such as "shape optimisation", to go further the classical use described by Izazaga-P\'erez \cite{Izazaga_Perez_FEM}. In the first section, we introduce rotational symmetry aspherics, known as conicoids, and the method to produce them using Active Optics and SMP more specifically by giving an example for a $3^{th}$-order spherical Zernike aberration. We will discuss the definition of shape optimization and the strategy adopted to implement it. Then, an algorithm has been developed allowing to optimize the deformation obtained, using data provided by Finite Element Model (FEM) outputs which has been implemented in Python language.

\section{Conicoid Mirror}
\subsection{Definition}
\label{sec:examples}
Primary mirrors are one of the most sensitive part of space telescopes. The increasing size of diameter allows collecting a large amount of photons and detecting fainter objects, while the surface quality is crucial to obtain the best images of stars, planets or exoplanets \cite{refSPHERE}. For these reasons, large diameter and high surface quality are requested on most space and ground-based future observatories. Nowadays most of the telescope optical designs use primary parabolic mirrors which are more convenient to focus rays from infinite objects without introducing any additional complexity. The sphere and the parabola are two peculiar solutions of what is commonly called \textit{conicoids}.

Let us define Eq.(\ref{eq:conicoids}) of a \textit{conicoids} aspheric optical surface with rotational symmetry by :

\begin{equation}
z = \frac{r^2}{R_c + \sqrt{R_c^2-(1+C)r^2} }
\label{eq:conicoids}
\end{equation}

where $R_c$ is radius of curvature of aspherics with a conic constant $C$ and $r$ is the radial coordinate from the center to the semi-diameter \textit{a}. Note that if $C=0$ the optical surface is a sphere and a parabola if $C=-1$.

By developing expression in expansion series, we obtain for the first terms Eq.(\ref{eq:expansion series}): 

\begin{equation}
z = \frac{r^2}{2R_c} + \frac{(1+C)r^4}{8R_c^3 } + \frac{(1+C)^2r^6}{16R_c^5 } + \cdots
\label{eq:expansion series}
\end{equation}

Retrieving the asphere $z_a$ from the sphere $z_s$ with radii of curvature $R_a$ and $R_s$ respectively allows us to determine the deformation $z_d$ to be applied during the polishing process as:

\begin{equation}
z_d = z_s - z_a
\label{eq:deformation}
\end{equation}

Replacing Eq.(\ref{eq:expansion series}) in Eq.(\ref{eq:deformation}) we obtain (Eq.(\ref{eq:deformation series})):

\begin{equation}
z_d = \frac{r^2}{2}(\frac{1}{R_s}-\frac{1}{R_a})  +  \frac{r^4}{8}(\frac{1+C}{R_s^3}-\frac{1}{R_a^3})  +  \frac{r^6}{16}(\frac{(1+C)^2}{R_s^5}-\frac{1}{R_a^5})  +  \cdots
\label{eq:deformation series}
\end{equation}

\begin{figure}[htpb]
 \centering
\begin{tikzpicture}[scale=1.]
\begin{axis}[axis lines=middle,
            xlabel=$r$,
            ylabel=$z$,
            enlargelimits,
            ytick={2.171,1.225},
            yticklabels={$z_s$,$z_a$},
            xtick={\empty}]
\pgfplotsset{compat=1.11}
\addplot[name path=F,black,domain={0:5}] {x^2*0.1} node[pos=.8, below,at end,yshift=-1cm,xshift=-0.3cm]{$ASPHERIC$};
\addplot[name path=G,black,thick,domain={0:5}] {(4.5-sqrt(4.5^2-x^2))+0.5}node[pos=.1, above,midway,yshift=2cm]{$SPHERE$};
\addplot[pattern=north east lines, pattern color=black!50]fill between[of=F and G, soft clip={domain=0:5}];
\node[anchor=east] at (axis cs:0,0.2) {$\epsilon$};
\filldraw[color=gray!50]
 (3.3,1.089) -- (3.3,1.94) 
 -- plot [domain=3.3:3.7] (\x,0.175*\x^2) 
 (3.7,2.44) -- (3.7,1.369) 
 -- plot [domain=3.7:3.3] (\x,0.1*\x^2) ;
 \draw[dashed,below left]node{$0$}
    {(0,2.171)--(3.5,2.171)node[above, midway]{$r$}}
    {(0,1.225)--(3.5,1.225)};
\draw[->](0,4)--(3.5,2.171)node[above, midway]{$R_s$};
\end{axis}
\end{tikzpicture}
\caption{Definitions of the best-fit sphere (BFS) minimizing the volume removal to obtain the targeted aspheric surface.}
\label{fig:Best-fit sphere}
\end{figure}
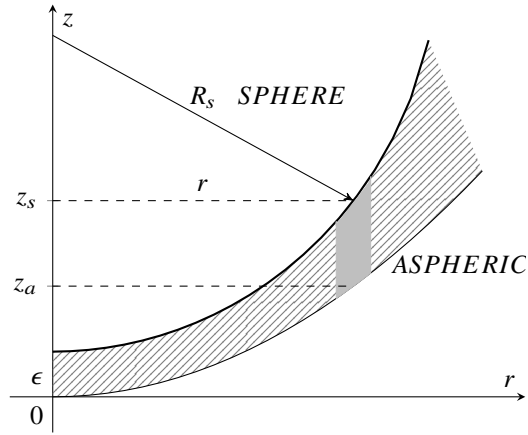

$R_s$ is defined as the first parameter of the substrate dimensions. $R_a$ is defined in order to minimize the removal volume defined by Eq.(\ref{eq:volume}) as described by Unti \cite{Unti:66}:

\begin{equation}
V=2\pi\int_0^r \Delta z r dr
\label{eq:volume}
\end{equation}

where $\Delta z$ is the difference in abscissa between the sphere and aspheric (Eq.(\ref{eq:Delta z})):

\begin{equation}
\Delta z = \epsilon + R_s - \sqrt{R_s^2-r^2} - z_a
\label{eq:Delta z}
\end{equation}

with $\epsilon$ is the shifted distance from the origin as shown in the Fig. \ref{fig:Best-fit sphere}.

\begin{equation}
\frac{dV}{dR} = 0
\label{eq:volume minimization}
\end{equation}

Unti \cite{Unti:66} demonstrated for a perfect parabolic mirror with a conic constant equal to -1, the best-sphere radius of curvature minimizing the volume removal (Eq.(\ref{eq:volume minimization})) is equal to: 

\begin{equation}
R_{BFS} = R_a + \frac{r^2}{4R_a}
\label{eq:radius of best sphere}
\end{equation}

The best-fit sphere radius can also be determined by minimizing the Root-Mean Square (RMS) deviation between the targeted optical surface and the best-fit sphere as used in the optical software ZEMAX \cite{Zemax2001}. The deformation $z_d$ is determined by substituting the radius of the sphere $R_s$ (Eq.(\ref{eq:radius of best sphere})) in the Eq.(\ref{eq:deformation}) by the best-fit sphere radius $R_{BFS}$.

\subsection{Active Optics and $3^{rd}$ order spherical aberration generation}

Stress mirror polishing developed by Lema\^itre \cite{Lemaitre:72} is derived from Schmidt works to aspherize his refractive correctors using full-size spherical tool to avoid any local retouch \cite{Everhart:66}. This technique relies on Active Optic methods using polishing pressure as an uniform load distribution. 

Active optics method based on elasticity's theory of materials is usually used to correct the Wave-Front Error (WFE) of an optical system by applying the adjusted deformation on one or several surfaces. Single or combined deformation modes can be generated in this purpose with one or several actuators. For example SPHERE, the Very Large Telescope (VLT) instrument dedicated to exoplanets detection, is assisted by this method as a means of compensating small deformations due to external constrains to the instrument, such as gravity load or temperature gradient, at long timescale \cite{procSPHERE}.

For years, they have also been deployed for many applications such as Variable Curvature Mirrors (VCMs) \cite{Lemaitre2009} implemented in the delay line system of the the Very Large Telescope Interferometer (VLTI) in order to compensate optical path differences by changing focii of mirrors also called zoom mirrors \cite{1998A&AS..128..221F}. Usually deformation modes are generated with a combination of pressure, forces and bending moments. However pure aspheric shapes are mostly difficult to execute on classical mirrors with quasi-constant thickness distribution. For this particular reason Variable Thickness Distribution (VTD) classes have been developed by Lema\^itre \cite{Lemaitre2009} to produce better first-order modes, by adapting the thickness distribution to the required deformation shape, the load distribution and the boundary conditions.

In our particular case, VTD is the starting point to generate aspherics using stress polishing. Third-order spherical aberration (SA3) can be generated by applying a uniform polishing pressure $P_p$ on the optical surface with a pushing central force $F_c$ in reaction, as developed by Lema\^itre \cite{Lemaitre2009_TSA3}, while the mirror's edge is maintained with a device discussed in the following paragraph. In that specific configuration, not only pure $3^{rd}$ order spherical aberration is produced but also curvature and some other high-order spherical aberrations which have to be removed.

The dimensionless thickness distribution $T_{SA3}^D$ which produces $3^{rd}$-order spherical aberration is described \cite{Lemaitre2009_TSA3} by the following Eq.(\ref{eq:T40}): 
\begin{equation}
T_{SA3}^D = [\frac{3+\nu}{1-\nu} \rho^\frac{-8}{3+\nu} - \frac{4}{1-\nu} \rho^{-2} + 1]^{1/3}
\label{eq:T40}
\end{equation}

with $\rho=r/a$\hspace{0.1cm} $\in[0;1]$ ; $\textit{a}$ the mirror's semi-diameter and $\nu$ the Poisson's ratio.

\begin{figure}[htbp]
\centering
\begin{tikzpicture}
\centering
\begin{axis}
[
xlabel={$Normalized \hspace{0.1cm} semi-diameter$},
ylabel={$T_{SA3}^D$},
xmin=0,
xmax=1.,
ymin=0,
ymax=18.,
grid=both,
grid style={line width=1.5pt, draw=gray!10}
]
\addplot[smooth,domain=0.05:1]{(((3+0.22)/(1-0.22) )*x^(-8/(3+0.22)) - (4/(1-0.22) )*x^(-2) + 1)^(1/3)};
\addplot[dashed,samples=200,domain=0.05:1]{x-x};
\end{axis}
\end{tikzpicture}
\caption{Dimensionless Variable Thickness Distribution $T_{SA3}^D$ in function of the normalized semi-diameter. This profile results from Eq.(\ref{eq:T40}).}
\label{fig:T40 distribution}
\end{figure}
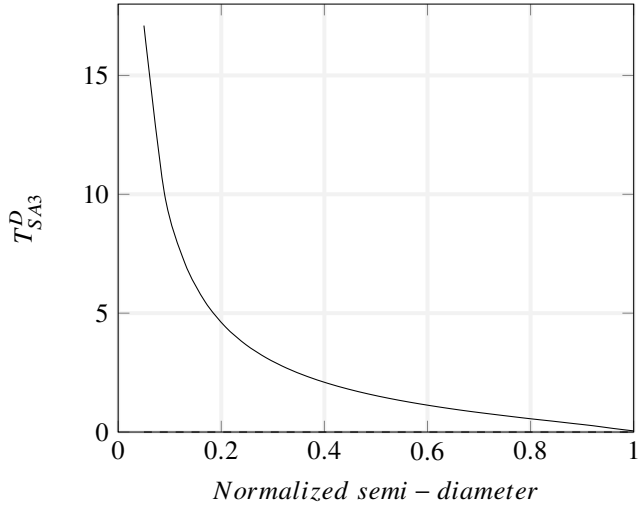

It is important to note that, in this solution, the thickness is zero at the mirror's edge and infinite at the center as shows in Fig. \ref{fig:T40 distribution}. In practice, primary mirrors have to be polished with an acceptable thickness at the edge according to the large mirror's diameter and finite thickness at the center.

To obtain the real variable thickness distribution $T_{SA3}$ (Eq.(\ref{eq:TSA3_real})), it is necessary to multiply the normalized dimensionless thickness distribution $T_{SA3}^D$ by a scaling factor $T_0$ \cite{Lemaitre2009_TSA3} and taking into account the sphere deviation and the edge thickness $e$. Scaling factor is depending on intrinsic parameters of the mirror's material such as Poisson's ratio $\nu$ and Young's modulus $E$ but also external loads $q$ and the sagitta of SA3 that expected to be generated $A_{SA3}$:

\begin{equation}
T_{SA3} = T_{SA3}^D*T_0 - (R_s - \sqrt{R_s^2-r^2}) + e
\label{eq:TSA3_real}
\end{equation}

with  \hspace{0.5cm} $T_0$ = $[\frac{3 (1-\nu^2) q}{16.A_{SA3}.E}]^{\frac{1}{3}}$

\section{Shape optimization and Pure $3^{rd}$-order spherical aberration}

Shape optimization is a very efficient tool as long as you know how to use it properly. Botkin \cite{Botkin} has defined the basis of such optimization to go beyond mass reduction problem using mainly size optimization of a model in which the geometry remains unchanged.  Shape optimization, has since been developed \cite{haftka} and is currently used in a large number of various applications such as fluid-structure interaction to find the optimal shape of a plane wing by minimizing drag \cite{mohammadi2010applied}. Very few examples are referenced in literature as noticed Lund \cite{LUND} for the fluid-structure coupling especially but the same could be said for field in optics. One should notice that it is not uncommon to encounter an amalgam between shape optimization and topology optimization. From an objective function, shape optimization refers to the relocation of points belonging to a given shape while topology optimization determines if a model element has to be removed or not based on the level of its \textit{involvement} in the model rigidity.

The shape optimization is based on the variation of design variables depending on an objective function, as mentioned before, which is usually a scalar subject to design constraints. The objective function can be a combination of design responses to provide a single final response. The main issue of using shape optimization is to generate the correct shape basis vectors. Indeed, each selected model gridpoints, also called \textit{nodes}, will move along these basis vectors directions. NASTRAN software integrates this functionality as described by Kodiyalam \cite{KODIYALAM1991821} but can also be generated by a self-programmed code. In this case, the model meshing should be adapted for both having enough space to relocate nodes and sufficient number \textit{N} of nodes to converge. Because of the large amount of design variables, the optimizer is not able to perform all the possible options but it can use design sensitivity analysis in order to determine the optimal set of design variables to reach the goal.

Design sensitivity computes all the rates of change \cite{Nastran2017}, also called \textit{partial derivatives} coefficients $\lambda_{ij}$, for each \textit{i}-th design variables $v_i$ to anticipate the influence on the \textit{j}-th model response $r_j$ given by Eq.(\ref{Partial derivatives}):

\begin{equation}
    \lambda_{ij}= \frac{\partial r_j}{\partial v_i}
\label{Partial derivatives}
\end{equation}

\subsection{Problem Formulation}

In our case we want to find the optimal variable thickness distribution with the aim of producing a pure $3^{rd}$-order spherical deformation defining by Zernike \textit{"Peak-to-Valley"}(PtV) polynomials \cite{Noll:76} as:

\vspace{-0.25cm}

\begin{equation}
z_{SA3}^{PtV} = (6\rho^4 - 6\rho^2)\hspace{0.1cm}z_{SA3}^{RMS}\hspace{0.1cm}\delta_{4,0}
\label{eq:SA3 theorique}
\end{equation}

with \hspace{0.2cm} $\delta_{n,m}$ = $\sqrt{n+1}$\hspace{0.2cm} X \hspace{0.2cm} 
$\left  \{
     \begin{array}{ll}
            \ 1 & \mbox{if} \hspace{0.2cm} m = 0 \\
          \ \sqrt{2} & \mbox{if} \hspace{0.2cm} m \neq 0
     \end{array}
    \right.$

\vspace{0.1cm}

$z_{SA3}^{RMS}$ is the root-mean square value of the $3^{rd}$-order spherical aberration linked to the PtV value using $\delta_{n,m}$ coefficient.

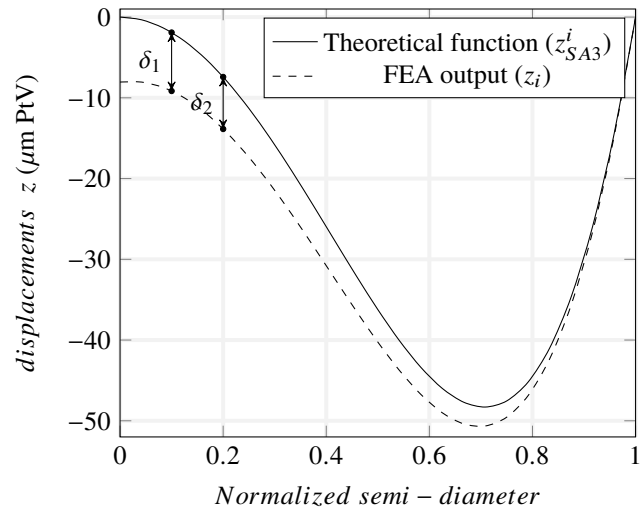
\begin{figure}[htbp]
    \centering
    \begin{tikzpicture}
    \centering
    \begin{axis}[
    xlabel={$Normalized \hspace{0.1cm} semi-diameter$},
    ylabel={$displacements \hspace{0.2cm} z$ ($\mu$m PtV)},
    xmin=0,
    xmax=1.,
    ymin=-52,
    ymax=1.,
    grid=both,
    grid style={line width=1.5pt, draw=gray!10}
    ]
    \filldraw[black] (20,381.41) circle (1pt);
    \filldraw[black] (20,445.81) circle (1pt);
    \filldraw[black] (10,428.42) circle (1pt);
    \filldraw[black] (10,500.87) circle (1pt);
    \draw[<->] (20,381.41) to node [left] {$\delta_2$} (20,445.81);
    \draw[<->] (10,428.42) to node [left]{$\delta_1$} (10,500.87);
    \plot[smooth,domain=0:1]{(6*x^4 - 6*x^2)*(14.4*sqrt(5))};
    \addlegendentry{Theoretical function ($z_{SA3}^i$)}
    \plot[dashed,smooth,domain=0:1]{(6*x^4 - 6*x^2 + (x-1)/4)*(14.4*sqrt(5))};
    \addlegendentry{FEA output ($z_i$)}
    \end{axis}
    \end{tikzpicture}
    \caption{Difference between theoretical function $z_{SA3}^i$ (solid line) and FEA output $z_i$ (dashed line)}
    \label{fig:Beta optimization}
\end{figure}

In this context, the goal of the optimization is to reduce the deviation of the deformation from the theoretical SA3 surface, given in Eq.(\ref{eq:SA3 theorique}), by minimizing the threshold $\beta(\Delta)$ with $\Delta = \delta_1,\delta_2,\dots,\delta_N$ and subject to the design constraints $\zeta_i$ (Eq.(\ref{Design constraints definition})):

\begin{equation}
    \zeta_i = \beta(\Delta)-\delta_i \geqslant 0 \hspace{0.5cm} \forall i \in [1,\ldots,N]
    \label{Design constraints definition}
\end{equation}

where $\delta_i$ (Eq.(\ref{eq:delta i})) is the deviation between model deformation responses $z_i$ and the SA3 theoretical shape for each gridpoints \textit{i} of the optical surface $z_{SA3}^i$, displayed in Fig. \ref{fig:Beta optimization}, defines as:

\begin{equation}
    \delta_i = z_i - z_{SA3}^i
    \label{eq:delta i}
\end{equation}

\begin{figure}[htbp]
    \centering
    \begin{tikzpicture}
    \centering
    \begin{axis}[
    xlabel={$Normalized \hspace{0.1cm} semi-diameter$},
    ylabel={$\delta_i$ ($\mu$m)},
    xmin=0,
    xmax=1.,
    ymin=0,
    ymax=10.,
    grid=both,
    grid style={line width=1., draw=gray!10}
    ]
    \filldraw[black] (20,62.79) circle (1pt) node [anchor=south west]{$\delta_2$};
    \filldraw[black] (10,72.45) circle (1pt) node [anchor=north]{$\delta_1$};
    \filldraw[black] (30,16.5*3.22) circle (1pt) node [anchor=south west]{$\delta_3$};
    \filldraw[black] (40,13.5*3.22) circle (1pt)node [anchor=south west]{$\delta_4$};
    \filldraw[black] (50,10.5*3.22) circle (1pt) node [anchor=south west]{$\delta_5$};
    \filldraw[black] (60,7.5*3.22) circle (1pt) node [anchor=south west]{$\delta_6$};
    \draw[<->] (10,72.45) to node [right] {$\zeta_1$} (10,82);
    \plot[thick,dash dot,samples=200,domain=0:1.1]{8.2};
    \addlegendentry{Objective function}
    \draw (25,90) node {$\beta(\Delta)$};
    \end{axis}
    \end{tikzpicture}
    \caption{Beta optimization principle. $\delta_i$ has to be decreased by reducing the objective function $\beta(\Delta)$}
    \label{fig:Beta optimization principle}
\end{figure}
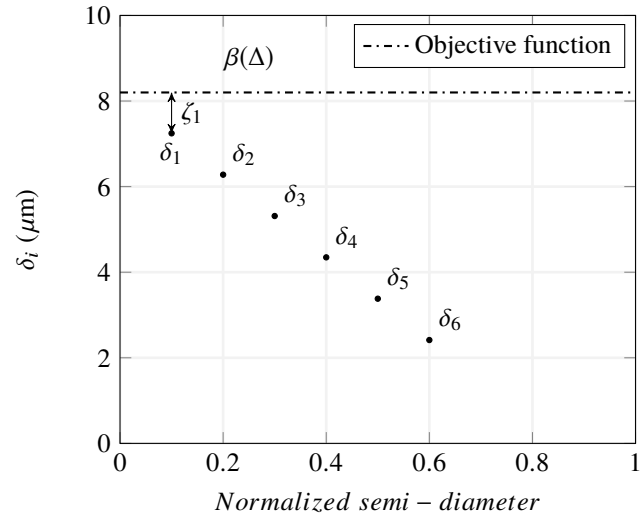

Fig. \ref{fig:Beta optimization principle} is showing the Beta optimization principle and the design constraints $\zeta_i$ which impose $\delta_i$ values to be lower than the threshold $\beta(\Delta)$ while it is reducing.

\newpage

\subsection{Shape optimization stategy}

\subsubsection{Shape optimization algorithm}
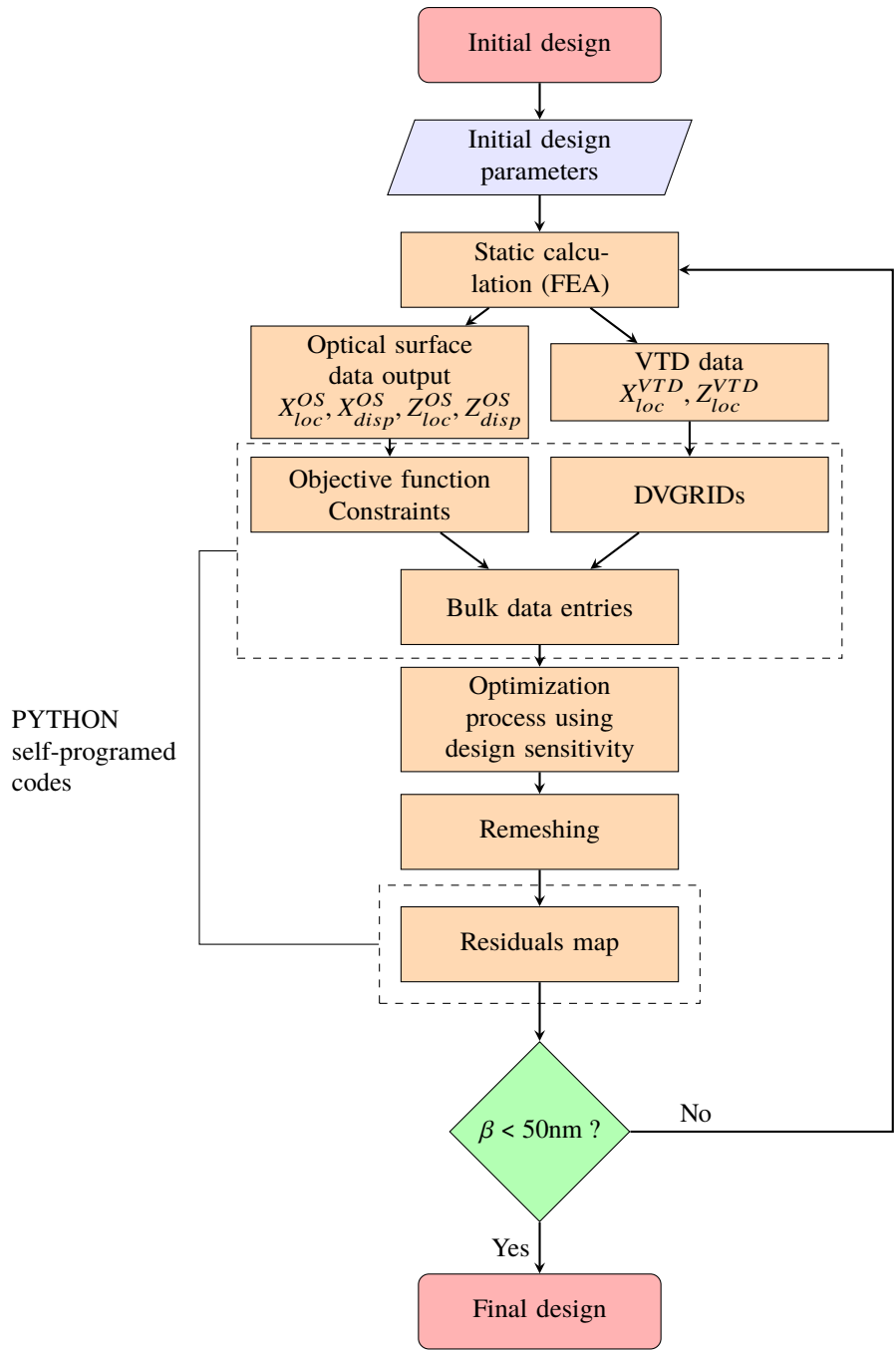
\begin{figure}[htbp]
\begin{tikzpicture}[node distance=1.5cm]
\tikzstyle{startstop} = [rectangle, rounded corners, minimum width=3cm, minimum height=1cm,text centered, text width=3cm, draw=black, fill=red!30]
\tikzstyle{io} = [trapezium, trapezium left angle=70, trapezium right angle=110, minimum width=2cm, minimum height=1cm, text centered, text width=3cm, draw=black, fill=blue!10]
\tikzstyle{process} = [rectangle, minimum width=3.7cm, minimum height=1cm, text centered,  text width=3cm ,draw=black, fill=orange!30]
\tikzstyle{decision} = [diamond, draw=black, fill=green!30]
\tikzstyle{arrow} = [thick,->,>=stealth]
    \node (start) [startstop] {Initial design};
    \node (parameters) [io, below of=start] {Initial design parameters};
    \draw [arrow] (start) -- (parameters);
    \node (static calculation) [process,below of=parameters] {Static calculation (FEA)};
    \draw [arrow] (parameters) -- (static calculation);
    \node (Optical surface data output) [process, below of=static calculation, xshift=-2cm] {Optical surface \\ data output \\ $X_{loc}^{OS},X_{disp}^{OS},Z_{loc}^{OS},Z_{disp}^{OS}$};
    \draw [arrow] (static calculation) -- (Optical surface data output);
    \node (VTD data output) [process, below of=static calculation, xshift=2cm] {VTD data $X_{loc}^{VTD},Z_{loc}^{VTD}$};
    \draw [arrow] (static calculation) -- (VTD data output);
        \node (Objective function) [process, below of=Optical surface data output] {Objective function \\ Constraints};
    \draw [arrow] (Optical surface data output) -- (Objective function);
    \node (DVGRIDs) [process, below of=VTD data output] {DVGRIDs};
    \draw [arrow] (VTD data output) -- (DVGRIDs);
    \node (Bulk data) [process, below of=static calculation, yshift=-3cm] {Bulk data entries};
    \draw [arrow] (Objective function) -- (Bulk data);
    \draw [arrow] (DVGRIDs) -- (Bulk data);
    \node (Optimization process) [process, below of=Bulk data] {Optimization process using design sensitivity};
    \draw [arrow] (Bulk data) -- (Optimization process);
    \node (Remeshing) [process, below of=Optimization process] {Remeshing};
    \draw [arrow] (Optimization process) -- (Remeshing);
    \node (Residual map) [process, below of=Remeshing] {Residuals map};
    \draw [arrow] (Remeshing) -- (Residual map);
    \node (Convergence) [decision,below of=Residual map, yshift=-1cm] {$\beta$ < 50nm ?};
    \draw [arrow] (Residual map) -- (Convergence);
    \node (fit) [draw,dashed,inner sep=5pt,fit={(Objective function) (DVGRIDs) (Bulk data)}]{};
    \draw [arrow] (Convergence.east) -- + (3.5,0) node [near start,above] {No} |-  (static calculation.east);
    \node (Final design) [startstop, below of=Convergence, yshift=-0.9cm] {Final design};
    \draw [arrow] (Convergence) -- node [anchor=east]{Yes}(Final design);
    \node (fit2) [draw,dashed,inner sep=8pt,fit={(Residual map)}]{};
    \draw [auto=left,thin,-] (fit.west) -- + (-0.5,0) node [at end,xshift=-1cm,yshift=-2cm,text width=3cm]{PYTHON \\ self-programed codes} |- (fit2.west);
\end{tikzpicture}
\centering
\caption{Flowchart of the optimization design}
\label{fig:Flowchart}
\end{figure}

Shape optimization method has to start with a model close to the optimal form such as the thickness distribution described in Eq.(\ref{eq:T40}) to avoid large gridpoints displacements and local minima. Finite Element Analysis (FEA) is operated for the first calculation with the initial model parameters to get the input data: optical surface displacements as design responses and variable thickness distribution as design variables layered in Fig. \ref{fig:Flowchart}.

Thereafter, the input data files are used to generate the bulk data entries with the aid of a self-programed PYTHON algorithm that will be embedded in the optimization input file. This program allows to calculate the maximum difference between the theoretical desired shape and the FEM displacements $\delta_{max} = Max (\delta_i)$ in order to define the minimum threshold to implement. Since the difference $\delta_{max}$ is small, a multiplying factor $c_3$ has to be applied to assure an objective function $\beta_{obj}$ large enough in order to prevent small values depending on the model unit (Eq.(\ref{eq:Beta objective})):

\begin{equation}
    \beta_{obj} = c_3*\beta(\Delta)
\label{eq:Beta objective}
\end{equation}

As mentioned, the threshold has to be higher than the maximum difference $\delta_{max}$ and has to be used as design constraint for each optical surface gridpoints. Eq.(\ref{eq:Design constraints}) describes the design constraint $\zeta_i$ implemented:

\begin{equation}
    \zeta_i = c_1*(c_2*\beta(\Delta) - |z_{SA3}-z_i|)
\label{eq:Design constraints}
\end{equation}

where $c_1$ and $c_2$ are the two coefficients used to manage the shape optimization. $c_2$ is slightly higher than the difference $\delta_{max}$ to avoid a constraint violation before running the optimization algorithm. The design constraint sensitivity is handled by the scaling factor $c_1$ depending on the design model.

As regards design variables selected on the thickness distribution boundary, they are all model gridpoints in this particular numerical simulation. They are compiled in a DVGRIDs entry defining the direction and magnitude for a given change in a design variable. 

Once the optimization is terminated, particularly for the first one, due to the deformation of the elements induced by the gridpoints displacements, it is indispensable to refine the Finite Element Model (FEM) to get more accurate results. With a projection on the Zernike orthonormal basis, the optical surface is decomposed into a linear combination of weighted orthogonal polynomials.

Due to the ill-posed problem such as non-adapted optimization coefficients or an initial design very far from the target, it is rare to converge in one step of optimization. In most cases, the optimization process needs a certain number of iterations, but usually less than a dozen. The optimization range will be performed on the entire mirror's surface $a$ but it can be narrowed to the interested area.

\subsubsection{Initial design}
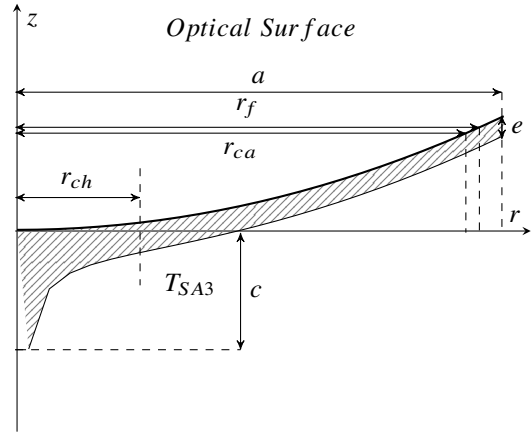
\begin{figure}
 \centering
\begin{tikzpicture}[scale=1.0]
\begin{axis}[axis lines=middle,
            xlabel=$r$,
            ylabel=$z$,
            xmin=-2,
            xmax=570,
            ytick={\empty},
            xtick={\empty}]
\addplot[name path=F,black,domain={-10:540},yscale=0.5,yshift=2.2cm] {-(((3+0.22)/(1-0.22) )*(x/570)^(-8/(3+0.22)) - (4/(1-0.22) )*(x/570)^(-2) + 1)^(1/3)+(3773.37-sqrt(3773.37^2-x^2))} node[pos=0.35, below,yshift=-0.3cm]{\textbf{$T_{SA3}$}};
\addplot[name path=G,black,thick,domain={0:540},yscale=0.5,yshift=2.7cm] {(3773.37-sqrt(3773.37^2-x^2))}node[,pos=.1, above,midway,yshift=2cm]{$Optical \hspace{0.1cm}Surface$};
\addplot[pattern=north east lines, pattern color=black!50]fill between[of=F and G, soft clip={domain=0:540}];
\draw[<->](0,580)--(540,580)node[above, midway]{$a$};
\draw[<->](0,520)--(515,520)node[above, midway]{$r_{f}$};
\draw[<->](0,510)--(500,510)node[below, midway]{$r_{ca}$};
\draw[<->](0,400)--(138.5,400)node[above, midway]{$r_{ch}$};
\draw[<->](540,540)--(540,500)node[right, midway]{$e$};
\draw[<->](250,340)--(250,140)node[right, midway]{$c$};
\draw[-](0,140)--(12,140);
\draw[dashed,below left]node{$0$}
    {(540,580)--(540,340)}
    {(515,520)--(515,340)}
    {(500,510)--(500,340)}
    {(0,140)--(250,140)}
    {(138.5,450)--(138.5,250)};
\end{axis}
\end{tikzpicture}
\caption{Initial 2D design with the optical surface (bold line) and the Variable Thickness Disctribution (solid line). The geometry parameters are discribed in the Table \ref{tab:Model parameters}.}
\label{fig:Initial design 2D}
\end{figure}

This first study is limited to a two-dimensional axisymmetric problem using Finite Element Analysis (FEA) software. In this approach, \textit{CTRIAX6} elements are required to make the calculation possible. However, \textit{CTRIAX6} element defines isoparametric and axisymmetric triangular elements with mid-side gridpoints which do not enable a large deformation due to the elements distorsion. Therefore, remeshing the model at each step is critical to pursue the optimization process. The initial design of the \textit{"tulip-form"} mirror based on the $T_{SA3}$ variable thickness distribution  is shown in Fig. \ref{fig:Initial design 2D}.

\newpage

The preliminary parameters introduced for the initial design are listed in Table \ref{tab:Model parameters}. We use a Zerodur vitro-ceramic substrate from Schott company as primary mirror material due to its low coefficient of thermal expansion perfectly adapted to space mirror stability. The mechanical properties of Zerodur are the following: Young's modulus= 91000 MPa, shear modulus=37295.08 MPa and density Poisson's ratio=0.22.

The outer mechanical diameter for SMP $a$ and the final outer mechanical diameter $r_f$ are 1080 mm and 1030 mm respectively as shown in the Figure \ref{fig:Initial design 2D}. The clear aperture has been chosen for more convenience and set at 1000 mm for a quasi-parabolic mirror with a conic constant around -0.967. The planned central hole is 277 mm diameter which will be machined after the polishing process. A starting mirror's edge thickness has been established at 2.9 mm in order to be able to support the mirror during the deformation. The total mass of the mirror before the central hole drilling is 25.29 kg.

\vspace{-0.1cm}

\begin{table}[htpb]
\centering
\caption{\bf Model parameters of the primary mirror}
\begin{tabular}{lc}
\hline
Model design parameters & Values \\
\hline
Initial outer mechanical diameter (for SMP) (a) & 1080 mm \\
Final outer mechanical diameter ($r_{f}$) & 1030 mm \\
Clear aperture diameter ($r_{ca}$) & 1000 mm \\
Initial Radius of curvature ($R_{s}$) & 3773.37 mm \\
Radius of the best-fit sphere ($R_{BFS}$) & 3792.69 mm \\
Central hole diameter added after polishing ($r_{ch}$) & 277 mm \\
Conic constant to achieve & -0.966682 \\
Edge thickness (e) & 2.9 mm \\
Central thickness (c) & 68 mm \\
Initial total mass & 25.29 kg \\
Final total mass with the central hole & 10.5 kg \\
\hline
\end{tabular}
  \label{tab:Model parameters}
\end{table}

\newpage

\subsubsection{Finite Element Model (FEM)}

A compromise between large gridpoint displacements and good accuracy in the first calculation was reached  with around 500 CTRIAX6 elements and 1200 gridpoints in the finite element model. These numbers should be increased when the objective function decreases in order to converge more effectively. 

\begin{figure}[htbp]
    \centering
    \includegraphics[width=0.8\textwidth]{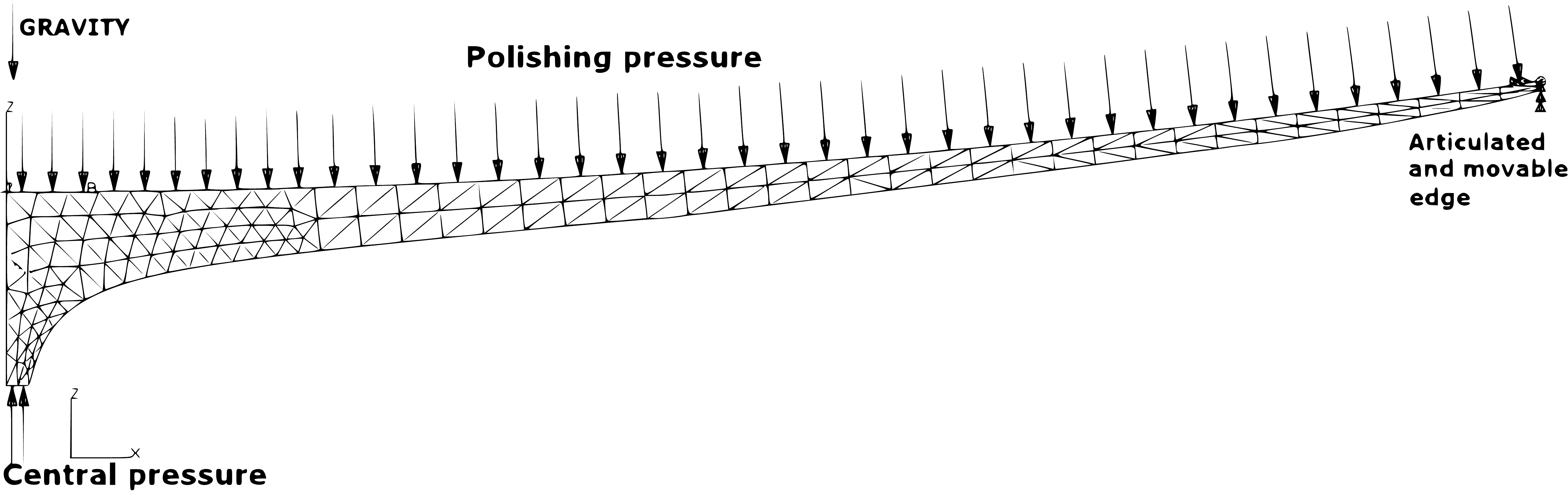}
    \caption{First meshing with initial design parameters and CTRIAX6 elements. Model loads are gravity, polisshing  and central pressures. An articuled and movable edge is implemented to increase the mirror's deflection.}
    \label{fig:Meshing}
\end{figure}

A combination of loads is applied in the FEA including a nominal polishing pressure of $31.5 g/cm^2$ on the optical surface, associated to a central pressure of $9.51 MPa$ at the \textit{"tulip-form"} basis to generate pure $3^{rd}$-order spherical aberration. Entire model is placed under $9.81 m.s^{-2}$ gravity as shown in Fig. \ref{fig:Meshing} and stays fixed during all the optimization process. 

An articulated and movable border along the radial axis is disposed at the top edge of the outer mechanical diameter. The purpose of this boundary condition is to avoid introducing radial moments $M_r$ and radial tensile forces $N_r$ and to allow a higher sagitta of mirror's deflection and a pure $3^{rd}$-order spherical aberration.

\subsubsection{Validity of the FEM}

Finite Element Model (FEM), and particularly Finite Element Analysis (FEA), is commonly used to study new opto-mechanical configuration before entering in the manufacturing phase of a mirror. In this approach the FEA outputs are cross-checked with some demonstrator experimental results in a second phase. Most of the time a very high correlation is noticed, however the last phase is mandatory to perfectly adjust the model. 

This procedure has been successfully used to design, simulate and polish toric mirrors of the SPHERE instrument for the VLT \cite{Hugot:08,Hugot_SMP_TM} or to modify the design of the warping harnesses in order to gain in performance on sky \cite{procSPHERE}. The same approach led to the study and development of active optics dedicated to space applications through the MADRAS deformable mirror \cite{Laslandes2011,Laslandes2013} and is currently applied for the design and manufacturing of off-axis parabola for the WFIRST mission \cite{Roulet2018}. 

This specific background in combining FEA with opto-mechanical fabrication, acquired during previous projects, give us a good confidence in the numerical results presented in the following section with respect to the up-coming prototyping phase.

\section{Results and Discussions}

The goal of the optimization is to deform the mirror by producing 14.4 $\mu$m RMS of $3^{rd}$-order spherical aberration during polishing process to obtain the targeted aspheric, which is an ellipsoid with a conic constant of -0.96682, for a full aperture diameter of 1080mm.
Optimization parameters, $c_1$, $c_2$ and $c_3$ coefficients have to be updated for each re-meshing step depending on design responses. DVGRIDs data entries are provided in function of design model but the  displacement of each gridpoints is decreasing as the final shape approaches. Because of the axisymmetrical model we are working on, we produce only axisymmetrical modes such as Focus and spherical aberrations. 

\begin{figure}
    \centering
    \includegraphics[width=0.65\textwidth]{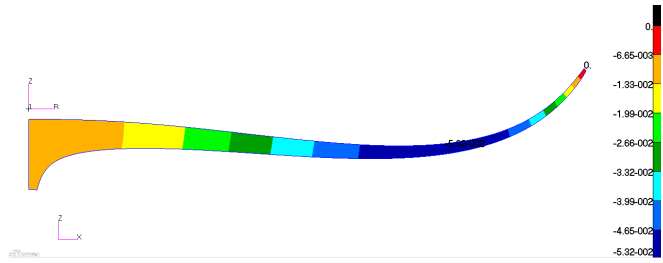}
    \caption{Deformation (in millimeters) with maximum (red) and minimum (blue) deflection values before the first optimization.}
    \label{fig:first deformation}
\end{figure}

Fig. \ref{fig:first deformation} shows the deformation resulting from the initial calculation before the first optimization. At first sight, the mirror has the shape of the planned $3^{rd}$-order spherical aberration. 

\begin{figure}
    \centering
    \includegraphics[width=0.8\textwidth]{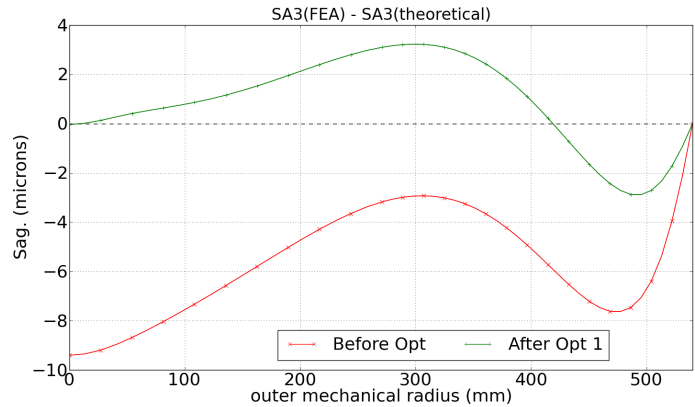}
    \caption{First optimization from the first deformation (figure \ref{fig:first deformation}) before optimization (bottom red curve) to the optimized deformation (top green curve). This curves represent the difference between FEA output and targeted theoretical aberration (in micrometers).}
    \label{fig:first optimisation}
\end{figure}

\begin{figure}
\centering
    \begin{tikzpicture}[scale=0.9]
    \centering
    \begin{axis}
    [
    xlabel={$Normalized \hspace{0.1cm} semi-diameter$},
    ylabel={$Normalized \hspace{0.1cm} sagitta$},
    xmin=0,
    xmax=1.,
    ymin=-1,
    ymax=1.,
    grid=both,
    grid style={line width=1.5pt, draw=gray!10},
    legend entries ={ Focus,
    SA3,
    SA5,
    },
    legend pos=south east
    ]
    \addplot[smooth,domain=0:1,blue,mark=square]{2*x^2-1};
    \addplot[smooth,domain=0:1,orange,mark=star]{6*x^4-6*x^2+1};
    \addplot[smooth,domain=0:1,black,mark=triangle]{20*x^6-30*x^4+12*x^2-1};
    \addplot[dashed,samples=200,domain=0.05:1]{x-x};
    \addlegendimage{line, blue}
    \addlegendimage{line, red}
    \addlegendimage{line, black}
    \end{axis}
    \end{tikzpicture}
    \caption{First three Zernike spherical polynomials: Focus (blue squares), SA3 (orange stars) and SA5 (black triangles).}
    \label{fig:First Zernike spherical polynomials}
\end{figure}
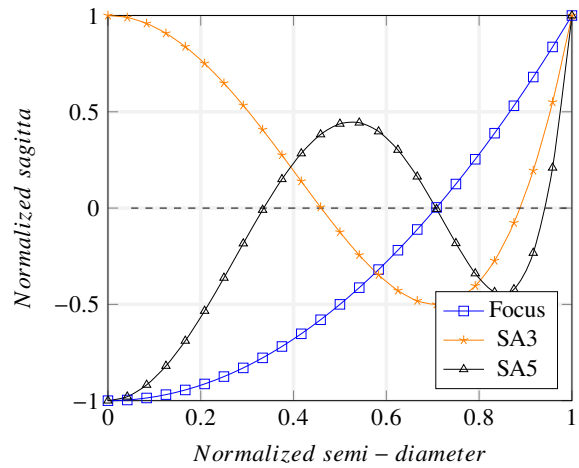

By subtracting the theoretical SA3 from the FEA deformation we can notice a large variation notably at the center of the mirror with 9.4 $\mu$m PtV surface error (red solid curve in Fig. \ref{fig:first optimisation}). The global deviation from the theoretical shape along the mirror's radius appears to be $5^{th}$-order spherical aberration (SA5) as shown in Fig. \ref{fig:First Zernike spherical polynomials}  depicting the three first theoretical Zernike spherical polynomials. The edge of the mirror is not moving along the optical axis, \textit{Z-axis}, due to the adopted boundary condition. The pupil is taking into account the radial displacement of the mirror in order order to have the right Zernike's decomposition.

\begin{figure}
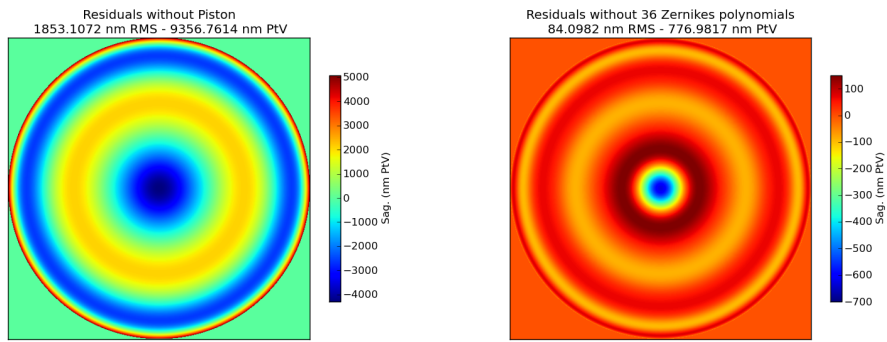
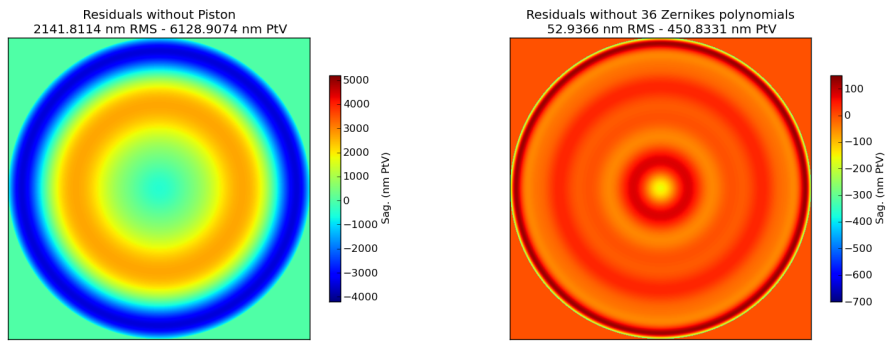

\centering
   \subfloat[Residuals without Piston\label{figur:1}]{%
    \includegraphics[width=0.5\textwidth]{Before_OPT_Residuals_without_Piston_SAG.pdf}}
    \hfill
   \subfloat[Residuals without 36 first Zernike polynomials\label{figur:2}]{%
     \includegraphics[width=0.5\textwidth]{Before_OPT_R36_SAG_Ech2.pdf}}
     \\
   \subfloat[Residuals without Piston\label{figur:3}]{%
    \includegraphics[width=0.5\textwidth]{After_OPT1_Residuals_without_Piston_0_540_SAG_Ech2.pdf}}
    \hfill
   \subfloat[Residuals without 36 first Zernike polynomials\label{figur:4}]{%
     \includegraphics[width=0.5\textwidth]{After_OPT1_R36_0_540_SAG_Ech2.pdf}}
    \caption{(a) residuals without Piston and (b) by retrieving numerically the 36 first Zernike polynomials (right) before the first optimization calculation on the outer mechanical diameter for SMP. Figures (c) and (d) show the results after the first optimization.}
    \label{fig:Residuals before and after first optimization}
\end{figure}

The initial calculation before the optimization process shows a deformation map without Piston (Fig. \ref{figur:1}) around 1.8 $\mu$m RMS, essentially SA5 with 1.4 $\mu$m RMS, as noticed previously. A projection of residuals on  Zernike modes between SA7 ($36^{th}$ Zernike polynomials) and $561^{th}$ Zernike polynomials indicates a high-order value of 84.1 nm RMS (Fig. \ref{figur:2}) with a maximum deflection at the center. Given the mirror's axisymmetrical geometry and the load condition, it is not necessary to use more than 561 Zernike polynomials to define the optical surface. Indeed, the spatial frequency between two peaks in the residual deformation is much lower in comparison with the radial component of the $561^{th}$ Zernike polynomial. 

\vspace{0.2cm}

A first optimization has been computed on the entire mechanical radius [0mm ; 540mm] letting gridpoints on the variable thickness distribution free to move along the Z-axis. A significant reduction of the difference between desired theoretical shape and FEA deformation can be noticed leading to a residual of 6 $\mu$m PtV (Fig. \ref{fig:first optimisation}) corresponding to a residuals sag. of 2.1 $\mu$m RMS on the entire pupil. Two extrema are present and contribute to the high-order spherical aberrations which can be visualised on the residuals error maps (Fig. \ref{fig:Residuals before and after first optimization}). 

The first error map (Fig. \ref{figur:3}) shows the residuals of the Zernike decomposition when the Piston aberration, which we do not need to consider, is substracted. This value decreases down to 52.9nm RMS by removing the 36 first Zernike polynomials by a numerical subtraction (Fig. \ref{figur:4}). These high-order polynomials need to be minimized as much as possible in the optimization process because of the complexity of applying the high-order removal procedure afterward.

\vspace{0.2cm}

\begin{figure}
    \centering
    \includegraphics[width=1.05\textwidth]{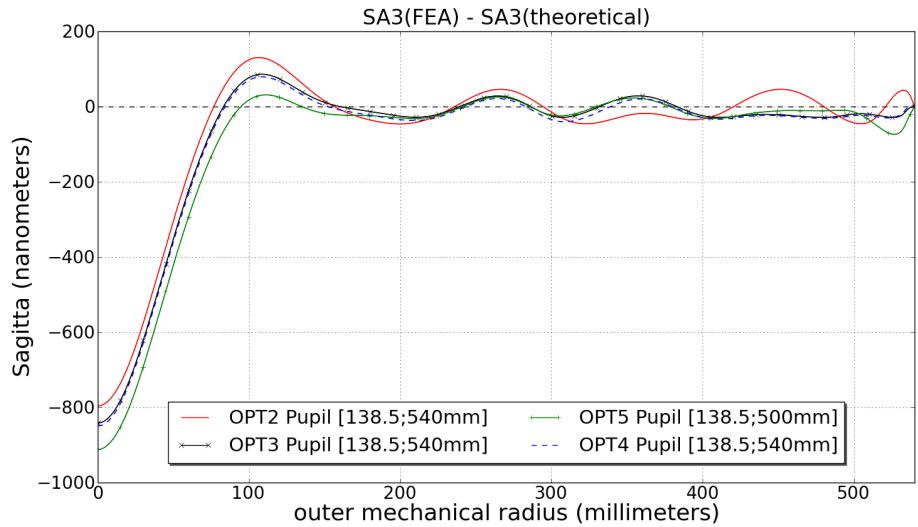}
    \caption{Next optimization steps of the optimization process performed on the reduced pupil [138.5;500mm]}
    \label{fig:Next optimization steps for the entire mirror's diameter}
\end{figure}

The next optimizations are performed on the reduced pupil [138.5mm ; 540mm] to take into account the central hole and thus release constraints on the optimization process. The deviation from the theoretical SA3 is free to evolve within the central hole while the optimization is focused on reducing the difference in the pupil of interest (Fig. \ref{fig:Next optimization steps for the entire mirror's diameter}).
As the threshold has to be decreased to reach a very low value of a few nanometers, the model unit has to be scaled in order to better control the optimization process. In this particular case, it was modified to use micrometers instead of millimeters to reach a better sensitivity on the optical surface coordinates.

\begin{figure}
    \centering
    \includegraphics[width=1.025\textwidth]{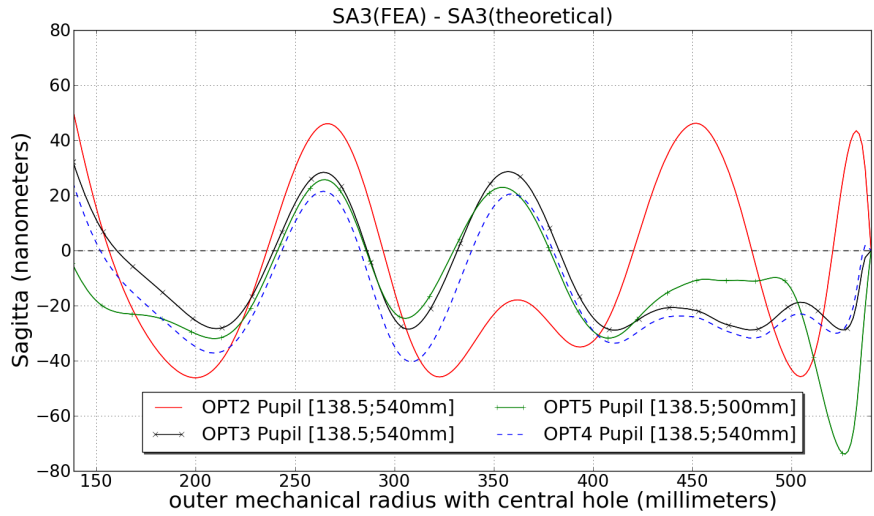}
    \caption{Zoomed view of the figure \ref{fig:Next optimization steps on the reduced pupil} on the area of interest [138.5;540mm]}
    \label{fig:Next optimization steps on the reduced pupil}
\end{figure}

By focusing the analysis on the reduced pupil it allows us to observe a substantial improvement in the aim of fitting the SA3 deformation given by the FEM model and the targeted theoretical SA3. 
However, some difficulties in the optimization process have to be noticed for a radius at distances larger than 400mm from the center. It is illustrated by a second regime (black and blue curves in Fig. \ref{fig:Next optimization steps on the reduced pupil}) retaining any optimization in this radius range. The residual error caps at around 20nm RMS on the drilled optical surface.  Indeed, this part of the mirror has a very small thickness, starting from 9.9mm to 2.9mm, which makes the model responses most sensitive to the design variables specially in this area. This leads to consider the optimization on the very limited pupil [138.5 ; 500mm] without the optical surface extension.

\vspace{0.2cm}

\begin{figure}
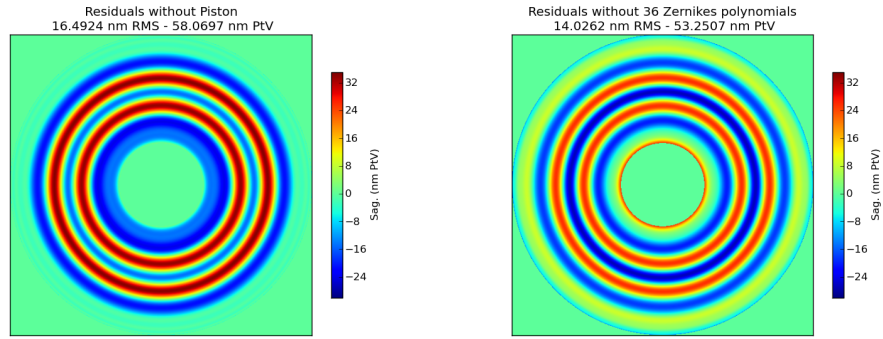

    \centering
   \subfloat[Residuals without Piston\label{figur:5}]{%
        \includegraphics[width=0.5\textwidth]{2D_After_OPT17_Residuals_without_Piston_SAG_Ech2.pdf}}
    \hfill
   \subfloat[Residuals without 36 first Zernike polynomials\label{figur:6}]{%
        \includegraphics[width=0.5\textwidth]{2D_After_OPT17_R36_SAG_Ech2.pdf}}
    \caption{(a) Residuals without Piston and (b) by retrieving numerically the 36 first Zernike polynomials (right) after the last optimization calculation on the clear aperture}
    \label{fig:Residuals last optimization}
\end{figure}

\begin{figure}
    \centering
    \includegraphics[width=0.6\textwidth]{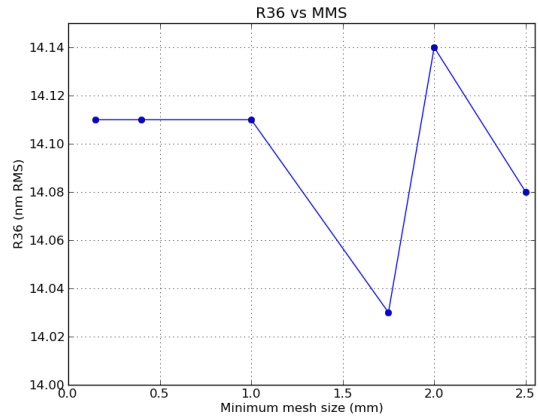}
    \caption{Evolution of the residuals without the 36 first Zernike polynomials (R36) with the central hole in function of the minimum mesh size (MMS)}
    \label{R36vsMMS}
\end{figure}

In this instance, a last optimization is performed on the final pupil [138.5;500mm] (green curve in Fig. \ref{fig:Next optimization steps on the reduced pupil}) adding a supplementary degree of freedom in the optimization for gridpoints located in the range [500;540mm]. Naturally, it leads to a peak in the non-optimized area to ensure a optimal result in the optimized one. Finally, the deformation map of the optical surface (Fig. \ref{figur:5}) shows a residual error, without piston, about 16.5 nm RMS in the reduced pupil including 14 nm RMS of high-order spherical modes after the 36 first Zernike's polynomials (Fig. \ref{figur:6}). After verification by changing the model units twice and refining it with very small elements (150 $\mu$m between two gridpoints), it was concluded that these residuals are effectively attributable to the FEA deformation and not to the accuracy of the FEA sampling (Fig. \ref{R36vsMMS}).

Indeed, by taking a minimum mesh size (MMS) (distance between two gridpoints) between 0.15mm and 2.5mm, the residual error without the 36 first Zernike polynomials (R36) ranges from 14.03 nm RMS to 14.14 nm RMS. This numerical difference of 1 Angstrom is comparable to the dimension of a single atom. In our case, a MMS of 1.75mm has been chosen to make a trade-off between precision of the results and reduction of computational time and load. In this manner, a duration of 30 minutes for each iteration in the optimization process has been performed with a total of five in our particular case.

\vspace{0.2cm}

\begin{figure}
    \centering
    \includegraphics[width=0.9\textwidth]{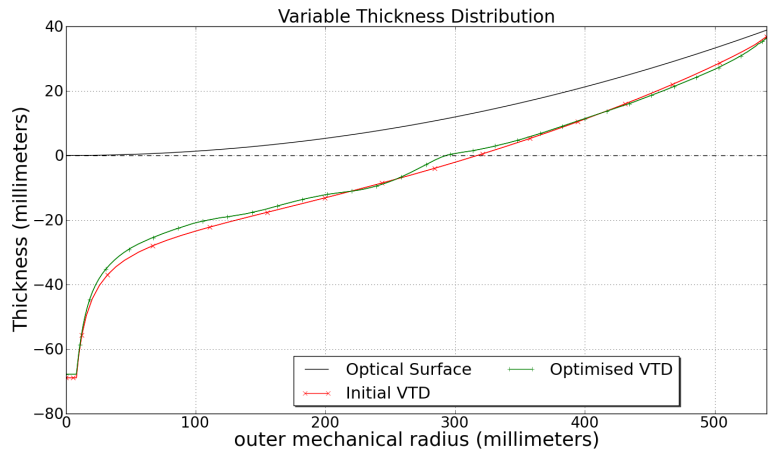}
    \caption{Initial (red curve) and optimised (green curve) Variable Thickness Distribution (VTD)}
    \label{fig:Optimised VTD}
\end{figure}

The optimized final variable thickness distribution in Fig. \ref{fig:Optimised VTD} (red curve) is different from the initial theoretical contour for many reasons. The first reason lies in the VTD definition which is valid for a zero thickness at the mirror's edge. In our case, the mirror needs to have a sufficient thickness at the edge to be maintained and a finite height at the center to respect manufacturing and loads specifications, leading to shift and cut the theoretical curve. Another reason explaining this VTD deviation is the useful surface used for the optimization. At the beginning, the entire optical surface has been exploited for the shape optimization but it has been reduced to the clear aperture to release some limitations and also accelerate the minimization process.

\vspace{0.2cm}

In addition, polishing and central pressures have been slightly adjusted to cancel focus and residual SA3 completely. In fact, the variation of spherical aberrations according to the loads is more sensitive for lower than higher Zernike modes. In this case, a minor modification of loads impacts very low orders leaving spherical residuals from SA5 not affected.

This adjustment can also be used, during manufacturing by SMP as an iterative process, to help converging towards former specifications. After the first polishing run, the optical part is usually slightly out of spec on very low orders. Adjusting polishing pressure and loads allow to compensate for over or under deformation of the optical part. A second and a third run are generally sufficient to reach the correct shape letting high spherical modes not impacted.

\vspace{0.2cm}

\begin{figure}
    \centering
    \includegraphics[width=0.8\textwidth]{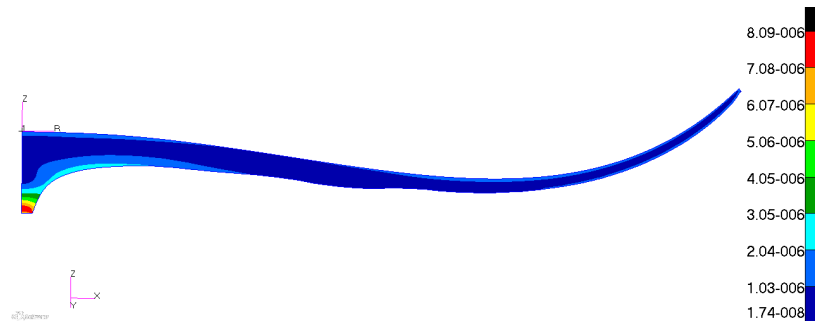}
    \caption{Maximum stress within the mirror (in $N.\mu m^{-2}$)}
    \label{fig:Maximum stress}
\end{figure}

The maximal Von Mises stress within the mirror during the deformation is equal to 8.09 MPa (Fig. \ref{fig:Maximum stress}) which is acceptable for standard use in this particular case (limited to 14 MPa). The maximal stress is located at the \textit{"tulip-form"} basis owing to the important pressure applied in a relatively small area. 

\section{Conclusions}

Thanks to this approach, combining SMP and shape optimization, we show a new manner to produce a conicoid, rotationaly symmetric aspheric, from a 2D finite element model by imposing pre-defined loads configuration and boundary conditions. The shape optimization method will assist the SMP process to reach high quality and very precise surface shapes in a minimum of iterations. In this study, DVGRIDs data entry have been defined in a basic way but they can be described with more degrees of freedom or even as a combination of orthogonal basis such as Zernike polynomials. It is also possible to define other theoretical symmetrical or non-symmetrical shapes to produce more complex optics with a combination of high-order modes. 

The main following step consists of polishing the thin mirror using the shape optimization method developed in this paper. The last step of the development will associate this large (>1m class) and thin (around 20mm) mirror to a rigid lightweight support in order to obtain very stiff reflective optics for space applications. In fact, sandwiches structures offer higher stiffness for lower mass ratio compared to other lightweighted designs \cite{Catanzaro}. In this case, the parallelization of the SMP technique using shape optimization and the rigid support manufacturing will participate to a considerable gain of fabrication time. Each part can be produced simultaneously on two different manufacturing procedures to be assembled in a final stage.

\section*{Funding}

Thales-SESO; Aix-Marseille Universite (AMU); Centre National de la Recherche Scientifique (CNRS).

\section*{Disclosures}

The authors declare that there are no conflicts of interest related to this article.


\bibliography{sample}






\end{document}